\documentclass[prd,showpacs,
a4paper, nofootinbib ,twocolumn,superscriptaddress]{revtex4}
\usepackage{graphicx}

\newcommand{\beq}{\begin{equation}}
\newcommand{\eeq}{\end{equation}}
\newcommand{\beqs}{\begin{eqnarray}}
\newcommand{\eeqs}{\end{eqnarray}}

\begin{document}
\title{Higgs boson from the meta-stable SUSY breaking sector}
\date{\today}

\author{Yang Bai \thanks{email: yang.bai@yale.edu}}
\affiliation{Department of Physics, Sloane Laboratory, Yale
University, New Haven, CT 06520}
\author{JiJi Fan \thanks{email: jiji.fan@yale.edu}}
\affiliation{Department of Physics, Sloane Laboratory, Yale
University, New Haven, CT 06520}
\author{Zhenyu Han \thanks{email: zhenyuhan@physics.ucdavis.edu}}
\affiliation{Department of Physics, University of California,
Davis, CA 95616}

\begin{abstract}

We construct a calculable model of electroweak symmetry breaking
in which  the Higgs doublet emerges from the meta-stable SUSY
breaking sector as a pseudo Nambu-Goldstone boson. The Higgs boson
mass is further protected by the little Higgs mechanism, and
naturally suppressed by a two-loop factor from the SUSY breaking
scale of 10 TeV. Gaugino and sfermion masses arise from standard
gauge mediation, but the Higgsino obtains a tree-level mass at the
SUSY breaking scale. At 1 TeV, aside from new gauge bosons and fermions similar to other little Higgs models and their superpartners, our model predicts additional electroweak triplets and doublets from the SUSY
breaking sector.

\end{abstract}


\maketitle

\section{Introduction}

Dynamical supersymmetry (SUSY) breaking has always been difficult
due to the existence of nonzero Witten index of supersymmetric
pure gauge theories~\cite{Witten:1982df}. It was believed before
that gauge theories  breaking SUSY dynamically must either be
chiral or have massless matter~\cite{Dynamical}. Models built
along this line turn out to be rather baroque. The situation is
greatly improved by the recent discovery of long-lived metastable
SUSY breaking vacua in $N=1$ supersymmetric QCD theory by
Intriligator, Seiberg and Shih (ISS)~\cite{Intriligator:2006dd}.
The ISS model opens up new possibilities to build realistic
models. Using ISS's idea, a number of new and simple models have
been constructed to communicate SUSY breaking (directly) to the
standard model~\cite{Directe Gauge Mediation}~\cite{R symmetry
breaking}.

In an interesting attempt, the ISS model is embedded into the
supersymmetric standard model by extending the Higgs
sector~\cite{Abel and Khoze}. But in that work,  a strong coupling
in the Higgs sector is needed to generate sufficiently large
gluino masses. In this paper we will construct an electroweak
symmetry breaking model with a weakly coupled Higgs sector
embedded inside the meta-stable SUSY breaking sector.

In our model, the soft masses of gauginos and sferminos are
generated by the standard gauge mediation. To obtain the soft
masses above  a few hundred GeV, the lowest allowed SUSY breaking
scale is  of order 10 TeV. We take two identical ISS sectors  with
$N_f=4$ massive fundamental flavors and set the SUSY breaking and
the global $U(4)^2$ breaking scale $\mu$ to be of order 10 TeV. We
then gauge the diagonal subgroup $SU(3)_W\times U(1)_X$ of the
unbroken global symmetry $U(3)^2$. By adding Yukawa couplings with
extra vector-like fermions, both of the $U(3)$ symmetries are
spontaneously broken at the scale  $f\sim O(\mu /4\pi)\sim
O(1~{\rm TeV})$. At the same time the  gauge symmetry is broken to
the electroweak symmetry. Among the resulting Goldstone boson
fields, one doublet is eaten by the heavy gauge bosons, while  the
other doublet is a pseudo Nambu-Goldstone boson (PNGB) and
identified as the Higgs doublet (this Higgs doublet is a light
linear combination of the two doublets $H_u$ and $H_d$ in  the minimal
supersymmetric standard model). The one-loop effective potential
of the Higgs doublet also breaks electroweak symmetry  at $v \sim
O(f /4 \pi)\sim O(\mu/(4\pi)^2) \sim O(100~{\rm GeV})$ and gives a
light Higgs boson with a mass from 100~GeV to 200~GeV. In short,
all the mass scales arise from one single scale $\mu$ in our
model. Since the scale $\mu$ can be  generated
dynamically~\cite{mu generating}, the hierarchy between the
electroweak scale and the Planck scale may also be explained.

Below the SUSY breaking scale, the ``collective symmetry breaking
mechanism" protects the Higgs boson mass as  in the little Higgs
models~\cite{little higgs}. Then in our model, like in the
super-little Higgs models~\cite{Roy:2005hg}~\cite{super little},
the Higgs boson mass is doubly protected by both SUSY and the
little Higgs mechanism, and the fine-tuning problem has been
alleviated  compared to the minimal supersymmetric standard model.
However, unlike the super-little Higgs models where the soft
masses are introduced by hand, all the soft masses in our model
are calculable. Our model contains new gauge bosons and fermions
at 1 TeV that are also present in other little Higgs models.
Furthermore, our model predicts additional electroweak triplets
and doublets at 1 TeV from the psedu-moduli in the meta-stable
SUSY breaking sector.

The paper is organized as follows. In Sec. II we will briefly
review the ISS model, its symmetries and its field content. In
Sec. III we present our model and describe how to break
$SU(3)_W\times U(1)_X$ to $SU(2)_w\times U(1)_Y$. In Sec. IV we
explain how to break the electroweak symmetry. We address the
vacuum alignment and order one quartic Higgs coupling issues. We
also explicitly minimize the full effective scalar potential to
support our step-by-step analysis. In Sec. V we discuss the soft
masses of gauginos and sferminos generated via gauge interactions.
In Sec. VI we provide a sample of the mass spectrum of our model.
We conclude in Sec. VII.

\section{Meta-stable SUSY breaking} \label{sec:ISS}

The ISS model is a deformed $N=1$ supersymmetric $SU(N_c)$ QCD,
with $N_f$ massive fundamental flavors. $N_f$ is taken to be in
the free magnetic range, $N_c+1\leq{N_f}<\frac{3}{2}N_c$ for a
controllable IR description of the theory. For concreteness, we
will concentrate on the simplest model $N_f=N_c+1$ where the
magnetic gauge group is trivial.

In terms of the superfields with normalized kinetic terms
\begin{center}
\begin{tabular}{c|ccc}
  & $SU(N_f)$ & $U(1)^\prime$ & $U(1)_R$ \\
 \hline
  $\psi$& $N_f$ & 1 & 0 \\
  $\tilde{\psi}$ & $\bar{N}_f$ & -1 & 0 \\
  $\Sigma$ & $Ad+1$ & 0 & 2 \\
\end{tabular}\,,
\end{center}
the low-energy effective superpotential we  study is
\begin{equation}
W=y(\tilde{\psi}\Sigma\psi-\mu^2{\rm Tr}\Sigma)\,,
\end{equation}
where $y$ is a dimensionless parameter of order one; $\mu$ is
related to a holomorphic scale $\Lambda$ of the microscopic
theory, but much smaller than $\Lambda$.  $\Sigma,
\psi,\tilde{\psi}$ are identified as mesons, baryons and
antibaryons of the electric theory. In addition, there is an
instanton generated operator $ det \Sigma/\Lambda^{N_{f}-3}$. For
$N_{f}>3$, this term is irrelevant and will be neglected in
discussions of the physics around the origin of $\Sigma$. Thus,
below we will set $N_f=4$. The reason for this choice will be
discussed later.

The F-terms of the meson field $\Sigma$ are
\begin{equation}
F_{\Sigma_{ij}}=y(\tilde{\psi}_{i}\psi_{j}-\mu^2\delta_{ij})\,.
\end{equation}
SUSY is broken since $\tilde{\psi}_{i}\psi_{j}$ has rank one while
$\delta_{ij}$ has rank $N_{f}=4$.  Up to global transformations,
the vacua are
\begin{equation}
\langle\Sigma\rangle= 0, \quad \langle\psi\rangle=\langle\tilde{\psi}\rangle^T=\left( \begin{array}{c} 0\\
\mu
\end{array}\right)\,.
\label{eq:vac}
\end{equation}
In these vacua, the global symmetry is broken to
$SU(3)\times{U(1)}\times{U(1)_R}$.

To see what the light fields are, we expand around
Eq.~(\ref{eq:vac}) using the following parametrization
\begin{equation}
\psi=\left( \begin{array}{c} H\\
\mu+\sigma\end{array}\right),
\tilde{\psi}^T=\left( \begin{array}{c} \tilde{H}^T\\
\mu+\tilde{\sigma}\end{array}\right), \Sigma=\left(
\begin{array}{cc} \Phi & N\\ \tilde{N} & Y
\end{array}\right),
\end{equation}
where the component fields transform under the unbroken global
symmetry as
\begin{center}
\begin{tabular}{c|cccccccc}

   & $H$ & $\tilde{H}$ & $\sigma$ & $\tilde{\sigma}$ & $\Phi$ & $N$ & $\tilde{N}$& $Y$ \\
  \hline
  $SU(3)$ & 3 & $\bar{3}$ & 1 & 1 & $Ad+1$ & 3 & $\bar{3}$ & 1 \\
  $U(1)$ & 1/3 & -1/3 & 0 & 0 & 0 & 1/3 & -1/3 & 0 \\
  $U(1)_R$& 0 & 0 & 0 & 0 & 2 & 2 & 2 & 2 \\
\end{tabular}
\end{center}
and ${\rm Tr}\Phi$ has a non-zero F-term  \beqs F_{{\rm
Tr}\Phi}\,=\,y\,\mu^2\,. \eeqs Another two fields, corresponding
to the Cartan generators in the adjoint representation, have
nonzero F-terms as well.

To identify the Goldstone fields, we use the following non-linear
parametrization of scalar fields \beqs
\psi&=&e^{\frac{\sigma_-}{\sqrt{2}\mu}}e^{
\frac{i}{\sqrt{2}\mu}\Pi_L}\, e^{ \frac{i}{\sqrt{2}\mu}\Pi_K}\,\left(%
\begin{array}{cc}
  0, & \mu+\frac{\sigma_+}{\sqrt{2}} \\
\end{array}%
\right)^T \,,\nonumber\\
\tilde{\psi}&=& \left(%
\begin{array}{cc}
  0, & \mu+\frac{\sigma_+}{\sqrt{2}} \\
\end{array}%
\right)\,e^{ \frac{i}{\sqrt{2}\mu}\Pi_K}\,
e^{-\frac{i}{\sqrt{2}\mu}\Pi_L}e^{-\frac{\sigma_-}{\sqrt{2}\mu}}  \,,
 \eeqs
 with
 \beqs
\Pi_L=\left(%
\begin{array}{cc}
  0 & L \\
  L^\dagger & 0 \\
\end{array}%
\right) \,,\quad \Pi_K=\left(%
\begin{array}{cc}
  0 & K\\
  K^\dagger & 0 \nonumber\\
\end{array}%
\right)\,.
 \eeqs
To the leading order, $L$ and $K$ are related to $H$, $\tilde{H}$ as
\beqs
L&=& \frac{1}{i\sqrt2}(H+\tilde{H}^\dagger)\,, \quad K=
\frac{1}{i\sqrt2}(H-\tilde{H}^\dagger)\,.
\eeqs
The scalar mass spectrum is
\begin{eqnarray}
m_{\sigma_{+}}^2&=&m_K^2\,=\,m_Y^2\,=\,2y^2\mu^2\,, \nonumber\\
m_{N}^2&=&m_{\tilde{N}}^2\,=\,y^2\mu^2\,,\nonumber \\
m_{\Phi}^2&=&m_{\sigma_{-}}^2\,=\,m_{L}^2\,=\,0\,,
\end{eqnarray}
where $K$ is the heavy combination of the triplets $H,\tilde{H}$,
while the massless combination $L$ together with the singlet $
Im(\sigma_{-})$ are the seven Nambu-Goldstone bosons (NGB)'s in
the coset space $SU(4)\times{U(1)^\prime}/SU(3)\times{U(1)}$. $Re
(\sigma_{-}), \Phi$ are pseudo moduli and obtain masses of order
$y^4\mu^2/(16\pi^2)$ through the one-loop Coleman-Weinberg
potential
\begin{equation}
V_{CW}^{(1)}=\frac{1}{64\pi^2}{\rm STr}
{\cal M}^4\log{\frac{{\cal M}^2}{\Lambda^2}}\,.
\end{equation}

In the UV regime, the instanton term in the superpotential can not
be neglected and is crucial to generate supersymetric vacua
$\langle{\Sigma}\rangle\sim{(\mu^2\Lambda)^{\frac{1}{3}}}$.  The
distance in field space between the local vacua and the global
minima is controlled by a small parameter $\mu/\Lambda$, and thus
the meta-stable vacuum can have a lifetime much longer than the
age of the Universe.

\section{The model}

The ISS model provides a way to break  SUSY  and can be used to
construct a gauge mediation model, where the soft masses of
gauginos and sfermions are roughly of the same scale,
$O(g^2_i\,F/(16\pi^2\,M))$. Here, $g_i\,(i=1,2,3)$ are the gauge
coupling constants in the standard model, $F$ is the value of an
F-term indicating SUSY breaking and $M$ is the supersymmetric mass
of the messenger. As the superpartner masses are of order 100~GeV
or more, we need to have $F/M \gtrsim 10$~TeV. To avoid tachyonic
directions of the messenger fields, $F< M^2$ and the lowest
allowed scale of $F$ is of order $(10~{\rm TeV})^2$.

Since  the Higgs boson mass is likely to be $O(100~{\rm GeV})$, it
has to be at least two-loop factor below the SUSY breaking scale
in the ISS model. Below the SUSY breaking scale, we  introduce the
collective symmetry breaking mechanism as in little Higgs models
to protect the mass of the Higgs boson. In the simplest little
Higgs model~\cite{Schmaltz:2004de}, the Higgs boson mass is
lighter than the $SU(3)$'s breaking scale $f\sim 1$~TeV by a
factor of $4\pi$. For our purpose, we need to achieve the
$SU(3)$'s breaking scale $f$ one loop factor lower than the SUSY
breaking scale $\sqrt{F}\sim 10$~TeV.

To achieve this purpose and to have enough light degrees of
freedom  containing the Higgs doublet as a PNGB, we adopt two ISS
sectors with $U(4)^2$ global symmetry. For simplicity, we choose
two identical ISS sectors by imposing a $Z_2$ symmetry between
them. In each ISS sector, the global symmetry $U(4)$ and SUSY are
broken at the same scale, $\sqrt{F}\sim \mu$, where the coupling
$y$ in the ISS sectors is chosen to be of order one. Then there
are two massless triplets $L_1$ and $L_2$, which are the NGB's in
each sector. Hence, the effective field theory below $\mu$
resembles the simplest little Higgs model with the unbroken global symmetry $U(3)^2$. After
gauging the diagonal $SU(3)_W\times U(1)_X$ subgroup, $L_1$ and
$L_2$  become PNGB's. We will later introduce a superpotential  to
spontaneously break  the approximate global $U(3)^2$ symmetry to
$U(2)^2$. The Higgs doublet is a PNGB and contained in the light
triplets as
 \beqs
L_1^T&=&f\left(i\frac{h}{|h|}\sin{\frac{|h|}{\sqrt{2}f}},\;\cos{\frac{|h|}{\sqrt{2}f}}\right)\,, \nonumber\\
L_2^T&=&f\left(-i\frac{h}{|h|}\sin{\frac{|h|}{\sqrt{2}f}},
\;\cos{\frac{|h|}{\sqrt{2}f}}\right)\,, \label{eq:para-triplets}
\eeqs where  $U(3)_1$ and $U(3)_2$ are broken at the same scale
$f$; $\sqrt{2}$ is chosen to have properly normalized kinetic term
of the Higgs doublet.

\subsection{Higgs sector and D-term}

For the two identical ISS sectors, the superpotential is \beqs
W&=&\;y\,({\rm Tr}\,\tilde{\psi}_1\Sigma_1\psi_1-\mu^2{\rm
Tr}\,\Sigma_1)\nonumber \\
&&+\,y\,({\rm Tr}\,\tilde{\psi}_2\Sigma_2\psi_2-\mu^2{\rm
Tr}\,\Sigma_2)\,. \eeqs The symmetry breaking and the vacua in both
sectors are the same as described in
section~\ref{sec:ISS}.

The parametrization around the ISS vacua are then \beqs
\psi_i=\left( \begin{array}{c} H_i\\ \mu+\sigma_i \end{array}
\right), \tilde{\psi}_i^T=\left( \begin{array}{c} \tilde{H_i}^T\\
\mu+\tilde{\sigma}_i \end{array} \right), \Sigma_i=\left(
\begin{array}{cc} \Phi_i & N_i\\ \tilde{N}_i & Y_i
\end{array}\right)\,,
\eeqs where $i=1,2$. The unbroken global symmetry is
$[SU(3)_1\times U(1)_1]\times [SU(3)_2\times U(1)_2]$. The
diagonal subgroup of this global symmetry is gauged and denoted as
$SU(3)_{W}\times U(1)_X$ with the electroweak gauge group as a
subgroup. Under the gauge symmetries, the field content in the
Higgs sector is described in Table~\ref{tab:higgs content}

\begin{table}[h!]
\begin{center}
\begin{tabular}{c|cccccccc}

   & $H_i$ & $\tilde{H}_i$ & $\sigma_i$ & $\tilde{\sigma}_i$ & $\Phi_i$ & $N_i$ & $\tilde{N}_i$& $Y_i$ \\
  \hline
  $SU(3)_W$ & 3 & $\bar{3}$ & 1 & 1 & $Ad+1$ & 3 & $\bar{3}$ & 1 \\
   $SU(3)_c$ & 1 & 1 & 1 & 1 & 1 & 1 & 1 & 1 \\
  $U(1)_{X}$ & 1/3 & -1/3 & 0 & 0 & 0 & 1/3 & -1/3 & 0
\end{tabular} \caption{Field content in the Higgs sector. $SU(3)_c$ is the  QCD group.} \label{tab:higgs content}
\end{center}
\end{table}

The D-term potential of $U(1)_X$ preserves the global
$SU(3)_1\times SU(3)_2$ symmetry, so it will not give a potential
to the Higgs doublet  and we neglect it here. To study the Higgs
doublet  potential, we ignore fields $N_i$, $\tilde{N}_i$ and
$\Phi_i$ and only consider the following part of the $SU(3)_W$
D-term potential
\begin{eqnarray}
\label{eq:dp} V_D&=&\frac{g^2}{2}\sum_{a}(\sum_{i}H_i^\dagger
t_aH_i-\tilde{H_i}^* t_a^*\tilde{H_i}^T)^2\,,
\nonumber \\
 &=&\frac{g^2}{2}\left((L_2^\dagger L_1)(K_1^\dagger K_2)
 +(K_1^\dagger L_2)(K_2^\dagger L_1)+{\rm h.c. }\right)
\nonumber \\
&&+\,\cdots\,,
\end{eqnarray}
where in the second line we expand the D-term potential in terms
of $L_i, K_i$, which are defined in section~\ref{sec:ISS}; $g$ is
the $SU(3)_W$ gauge coupling; $t_a$ are the gererators of
$SU(3)_W$ and ${\rm tr}[t^a,t^b]=\frac{1}{2}\delta^{ab}$; the dots
represent other $SU(3)_1\times SU(3)_2$ preserving terms and do
not contain quartic terms of $L_1$ and $L_2$. Substituting
$\langle K_i\rangle=0$, the D-term does not provide a
self-interaction potential of $L_i$ and the Higgs doublet at tree
level (the Higgs doublet is embedded in $L_i$ as in
Eq.~(\ref{eq:para-triplets})). Actually, this result comes from
the same vacuum expectation values (VEV)'s of $\psi_i$ and
$\tilde{\psi}_i$ in the ISS model. After integrating out the heavy
modes, the one-loop effective potential of the two light triplets
from the gauge interaction is of the form
\begin{eqnarray}
V&=&O(\frac{g^2}{16\pi^2})(\mu^2|L_1|^2+\mu^2|L_2|^2+|L_1|^4+|L_2|^4
\nonumber \\
&&+\,|L_1|^2|L_2|^2+|L_1^\dagger
L_2|^2)+\,...\,,
\end{eqnarray}
where order one numbers are neglected in the parenthesis.
Substituting the parameterization of $L_1$ and $L_2$ in
Eq.~(\ref{eq:para-triplets}) into this potential,  the Higgs
doublet  mass is $O(g/4\pi)f$ and it is at most of order 100~GeV.

\subsection{Breaking $SU(3)_W\times U(1)_X$ to $SU(2)_w\times U(1)_Y$}

To spontaneously break the global $SU(3)$ symmetries, we employ the trick of adding vector-like fermions
that has been widely used in the little Higgs model building. The
superpotential we propose is
\begin{eqnarray}
\label{eq:sp}
&\sum_{i=1}^2&[y_1\,\,Q_i\,\psi_i\,p_i^c+y_2{\rm
Tr}\,\Phi_i\,T_iT_i^c+y_3\,(\mu\,T_iX_i^c +\mu\,T_i^cX_i)\nonumber \\
&&+y_4\,\mu\, T_iT_i^c]\,,
\end{eqnarray}
where the superfields $Q_1\equiv({\cal Q},T_1)$ and $Q_2\equiv({\cal
Q},T_2)$ with ${\cal Q}\equiv(t,b,p)$. Here $t$ and $b$ are top and bottom left-handed quarks. The charge assignments of those new fields are listed in Table~\ref{tab:addtional-fields}.
\begin{table}[h!]
\begin{center}
\begin{tabular}{c|cccccc}
  & ${\cal Q}$ & $T_i$ &  $T_i^c$ &  $p_i^c$ &  $X_i$ &  $X_i^c$  \\
  \hline
  $SU(3)_W$ & $\bar{3}$ & 1 & 1 & 1 & 1 & 1  \\
  $SU(3)_c$ & 3 & 3 & $\bar{3}$
    & $\bar{3}$  & 3  & $\bar{3}$ \\
  $U(1)_X$ & 1/3 & 1/3 &  -1/3 &  -2/3 &  1/3 &  -1/3  \\
\end{tabular}
\caption{Field content in the top quark sector.}
\label{tab:addtional-fields}
\end{center}
\end{table}
We only list  fields in the top sector, which provides the
dominate contribution to the Higgs doublet mass. The full list of
fields without gauge anomalies can be found in~\cite{Roy:2005hg}.

In order not to change the ``rank condition" of the ISS model and
not to generate VEV's for the $SU(3)_c$ charged fields, the
following relation among Yukawa couplings must be satisfied
  \beqs
y^2_2\,F^2_{{\rm Tr}\Phi_i} < y_3^4\,\mu^4\Leftrightarrow
\,y^2y_2^2 < y_3^4\,. \label{eq:yukawa-condition}
  \eeqs

Other than the superpotential in Eq.~(\ref{eq:sp}), we also need an
additional superpotential to generate order one quartic couplings
for $L_1$ and $L_2$. We introduce extra singlets $Z$'s which
couple to $H_i$ and $\tilde{H}_i$ as
\begin{equation}\label{eq:sp-quartic-triplet}
y_5\,(Z_1\tilde{H}_1H_1+Z_2\tilde{H}_2H_2)\,,
\end{equation}
where $y_5$ is of order one. This superpotential is $U(3)^2$
invariant and contains operators $|L_1|^4$ and $|L_2|^4$ at
tree level. Additional operators like $|L_i|^2|K_i|^2$ from this superpotential do not provide a self-interaction potential of $L_i$ at tree level, but they contribute to the masses of $L_i$ at one-loop level. We will neglect their contributions by choosing $y_5$ to be smaller than the Yukawa couplings in Eq.~(\ref{eq:sp}).

\begin{figure}[h]
\begin{center}
\includegraphics[width=1\linewidth]{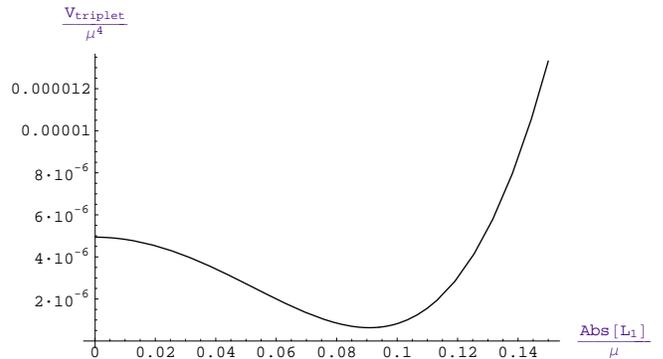}
\caption{Plot of $V_{\rm triplets}(L_1,L_2=0)$ as a function of
$L_1$. Here, we have subtracted  the potential by an $O(\mu^4)$
constant. This plot  shows that
the triplet $|L_1|$ develops a nonzero VEV at $0.09\mu$, and breaks  $U(3)_1$ to
$U(2)_1$. \label{fig:triplet-potential}}
\end{center}
\end{figure}

From Eq.~(\ref{eq:sp}), we  calculate the one-loop
Coleman-Weinberg potential for those light triplets $L_1$ and
$L_2$. Together with the order one quartic couplings,  the
potential of those two triplets is
  \beqs
V_{\rm triplets}(L_1,L_2)=V_{CW}(L_1,L_2)+
\frac{y^2_5|L_1|^4}{4}+\frac{y^2_5|L_2|^4}{4}\,. \label{eq:quartic-triplet}
 \eeqs
The fermion and scalar mass matrices  are too complicated
to be diagonalized analytically, and hence we present a
numerical study here. Choosing $y=y_2=y_4=1$, $y_1=y_3=2$ (satisfying the condition in
Eq.~(\ref{eq:yukawa-condition})) and $y_5=0.5$, we plot the triplet potential along the $L_1$ direction in
Fig.~\ref{fig:triplet-potential}.
Fig.~\ref{fig:triplet-potential} shows that
the triplet $|L_1|$ develops a nonzero VEV at $0.09\mu$, which
parametrically is \beq \langle |L_1| \rangle \,=\,
f\,=\,O(\mu/4\pi)\,,\eeq and breaks  $U(3)_1$ to
$U(2)_1$. A similar result can be obtained for
$L_2$ to break $U(3)_2$ to $U(2)_2$.

A few comments are in order about the superpotential in
Eq.~(\ref{eq:sp}) and Fig.~\ref{fig:triplet-potential}:

\begin{itemize}

\item The scale $O(10^{-6})$ on the vertical axis of
Fig.~\ref{fig:triplet-potential} comes from the product of the
one-loop suppression factor and small values of $\langle
L_1\rangle$ near the minimum.

\item The potential of $L_i$ is periodic as $L_i$ are PNGB's. It is
found that the minimum in Fig.~\ref{fig:triplet-potential} is the
global one, so we only consider the parameter space near the
origin.

\item If the operator $L_1^\dagger L_2$ were generated in our
effective potential, naive dimensional analysis suggests it would
give an $O(f^2)$ contribution to the Higgs doublet mass. However,
our superpotential preserves the $U(1)_1\times U(1)_2$ symmetry
which forbids the  $L_1^\dagger L_2$ term.

\item  The gauge couplings are smaller than the Yukawa couplings in Eq.~(\ref{eq:sp}), so we neglect the contributions from the gauge interactions to the potential of $L_i$ in Fig.~\ref{fig:triplet-potential}.

\item Logarithmic divergence for the one loop potential of the
triplets from Eq.~(\ref{eq:sp}) is absent. This is due to the
double protection of SUSY and the collective symmetry breaking
mechanism on the masses of PNGB's or triplets here. Without SUSY,
more than one Yukawa coupling is needed to generate a potential
for PNGB's, and then there is not any quardratic divergent
potential for the PNGB's at one loop. SUSY protects the mass of
PNGB's furthermore and leaves us one finite potential for those
triplets at one loop.

\end{itemize}

 After $L_1$ and $L_2$ develop nonzero VEV's as $L^T_1=L^T_2=f\,(0\,,0\,,1)$
 (the alignment issue will be discussed later), the global symmetry $U(3)^2$
 spontaneously breaks to $U(2)^2$ and generates two doublets as NGB's. The gauge
 symmetry $SU(3)_W\times U(1)_X$ spontaneously breaks to $SU(2)_w\times U(1)_Y$.
 One of those two NGB doublets is eaten by the gauge bosons, $W^\prime$'s and $Z^\prime$,
 and the other one becomes a PNGB, which is the Higgs doublet in our model and
 parameterized in Eq.~(\ref{eq:para-triplets}). The D-term also gives an additional
 mass to one linear combination of  $K_1$ and $K_2$ to complete the ``super Higgs mechanism".

\section{Electroweak symmetry breaking}

Other than providing negative mass terms to $L_1$ and $L_2$, the
one-loop Coleman-Weinberg potential of $L_i$ calculated from Eq.~(\ref{eq:sp}) also contains an operator,
$|L_1L^\dagger_2|^2$, with a positive $O(1/(16\pi^2))$
coefficient. For example, using the same values of Yukawa
couplings as in the previous section, the coefficient is
$+\,0.5/(16\pi^2)$. Substituting the parametrization of $L_1$ and
$L_2$ in terms of the Higgs doublet in Eq.~(\ref{eq:para-triplets}), we have
 \beqs
V_{CW}(L_1,L_2)&=&O(\frac{1}{16\pi^2})|L_1L^\dagger_2|^2\,+\,\cdots\,,
\nonumber \\
&=&-\,O(\frac{f^2}{16\pi^2})\,h\,h^\dagger\,+\,\cdots\,.
 \eeqs
Here $L_1L_2^\dagger = f^2 -  hh^\dagger+\cdots$. The Higgs doublet mass is negative and the
electroweak symmetry  successively breaks to $U(1)_{em}$.
However, this result is based on the assumption that the VEV's of
$L_1$ and $L_2$ are aligned (the VEV's of both triplets have only
nonzero values at the third entries).  Therefore, we need to address
the vacuum alignment issue first.

\subsection{Vacuum alignment}

Although the operator $|L_1L^\dagger_2|^2$ with a positive
coefficient gives a negative mass to the Higgs doublet, it prefers
anti-aligned VEV's of $L_1$ and $L_2$. Thus, we need to introduce
a new operator together with $|L_1L^\dagger_2|^2$ to achieve the
alignment and a negative mass for the Higgs doublet at the same
time. One possibility is to include an operator, $y^\prime
Z_3(\tilde{H}_1H_2-\mu^2)$, which is the only operator explicitly
breaking one linear combination of $U(1)_1\times U(1)_2$.
Therefore, the coefficient $y^\prime$ can  be as small as possible
and we choose it to be $O(1/16\pi^2)$. Combining the $F$-term
potential of $Z_3$ and $O(1/16\pi^2)|L_1L^\dagger_2|^2$, we have
$SU(3)_1\times SU(3)_2$ breaking parts of the potential as
 \beqs
 V\,=\,-\,v^2_1(L_1L_2^\dagger\,+\,L_2L_1^\dagger)/2\,+
 \,\frac{v^2_2}{f^2}|L_1L^\dagger_2|^2\,+\cdots\,,
 \label{eq:vac-align}
 \eeqs
where we define $v_1\equiv y^{\prime} \mu = O(1/16\pi^2)
\mu =O(1/4\pi)f$ and the positive $O(1/16\pi^2)$ coefficient in
front of $|L_1L^\dagger_2|^2$ as $v^2_2/f^2$. For this potential,
under the following condition
 \beqs
v^2_2 < v^2_1 < 2v^2_2 \,,
 \eeqs
the VEV's of $L_1$ and $L_2$ are aligned and at the same time the
Higgs doublet has a negative mass
 \beqs
-\frac{1}{2}(4v^2_2 \,-\,2v^2_1)h\,h^\dagger \,.\label{eq:higgs-mass}
 \eeqs

\subsection{Order one quartic Higgs coupling}
To obtain the electroweak symmetry breaking scale at 246 GeV, we
need to introduce an order one quartic Higgs coupling, which is
difficult without generating a large Higgs mass term. This problem
is present not only in a realistic super little Higgs model but
also in the simplest little Higgs model. In our model, we use the
``sliding singlet mechanism"~\cite{slide} to reach this goal by
including the following renormalizable superpotential:
 \beqs
&&y_6[Z_4(S_1^2\,-\,\tilde{H}_1H_1)\,+\,Z_5(S_2^2\,-\,\tilde{H}_2H_2)
\nonumber\\
&+&Z_6(S_3S_4\,-\,\tilde{H}_1H_1)\,+\,Z_7(S_3S_4\,-\,\tilde{H}_2H_2)
\nonumber \\
&+&Z_8(S_1S_3\,-\,\tilde{H}_1H_2)\,+\,Z_9(S_2S_4\,-\,\tilde{H}_2H_1)]\,.
 \eeqs
 By assigning appropriate charges to $Z_i$ and
$S_i$, this superpotential also preserves the global $U(1)_1\times
U(1)_2$ symmetry. The $F$-term potential of the $Z_i$'s is
 \beqs
&&y_6^2(|S_1^2\,-\,L_1L^\dagger_1/2|^2\,+\,|S_2^2\,-\,L_2L^\dagger_2/2|^2
\nonumber\\
&+&|S_3S_4\,-\,L_1L^\dagger_1/2|^2\,+\,|S_3S_4\,-\,L_2L^\dagger_2/2|^2
\nonumber \\
&+&|S_1S_3\,-\,L_2L^\dagger_1/2|^2\,+\,|S_2S_4\,-\,L_1L^\dagger_2/2|^2)\,.
\label{eq:quartic-doublet}
 \eeqs

Minimizing this potential, one can see that the VEV's of $S_i$
cancel the VEV's of $L_i$. After substituting the parametrization
of $L_i$ in Eq.~(\ref{eq:para-triplets}),  we have a potential of
the Higgs doublet without any VEV's. Hence, the quartic coupling
of the Higgs doublet is of order one, and with a negative mass
from Eq.~(\ref{eq:higgs-mass}) the Higgs doublet develops a VEV at
$O(v) = O(1/4\pi)f$.

\subsection{Complete triplet potential}
Up to now, we have first studied the spontaneously  breaking of
$SU(3)_W\times U(1)_X$ to $SU(2)_w\times U(1)_Y$, and then studied
the Higgs doublet potential to break the electroweak symmetry. On
the other hand, we can also study the full $SU(3)_W\times U(1)_X$
invariant potential of the two light triplets $L_i$, minimize
their potential and derive the final VEV structure. Combining
Eqs.~(\ref{eq:quartic-triplet}),~(\ref{eq:vac-align}),~(\ref{eq:quartic-doublet})
and choosing $y_5=y_6=1/2$ (the reason to choose a smaller $y_6$
than the Yukawa couplings in Eq.~(\ref{eq:sp}) is the same as
$y_5$), we have the following complete potential
\beqs
V&=&\frac{(|L_1|^2\,-\,f^2)^2}{16}\,+\,\frac{(|L_2|\,-\,f^2)^2}{16}-\,\frac{v^2_1(L_1L_2^\dagger\,+\,L_2L_1^\dagger)}{2}
\nonumber \\
&&+\frac{v^2_2}{f^2}|L_1L_2^\dagger|^2+(|S_1^2\,-\,|L_1|^2/2|^2+|S_2^2-|L_2|^2/2|^2
\nonumber \\
&&+|S_3S_4-|L_1|^2/2|^2+|S_3S_4-|L_2|^2/2|^2
\nonumber \\
&&+|S_1S_3-L_2L^\dagger_1/2|^2+|S_2S_4-L_1L^\dagger_2/2|^2)/4\,,
 \eeqs
with $f=O(1/4\pi)\mu$ and $v_1,v_2=O(1/4\pi)f$ determined by the
Yukawa couplings in Eq.~(\ref{eq:sp}). The  crucial ``--" signs in
the first two terms and the ``+" sign before the fourth term are
also derived from Eq.~(\ref{eq:sp}). Here, we neglect the
contributions to the effective potential from the gauge
interaction, due to the smallness of the gauge coupling compared
to the Yukawa couplings. For simplicity, we neglect the quartic
operator $|L_1|^2|L_2|^2$ generated at one loop, which does not
change our final result significantly.

Minimizing this potential, we derive
 \beqs
\langle |S_i|\rangle &=&
\sqrt{\frac{f^2\,+\,8v^2_2\,+\,6v^2_1}{2\,+\,40v^2_2/f^2}}\,, \nonumber \\
\langle L_1\rangle^T&=&  f^\prime\left(%
\begin{array}{ccc}
  i\,\sin{\frac{\langle h\rangle}{ \sqrt{2}f^\prime}}\,, &
  0\,, &
  \cos{\frac{\langle h\rangle}{ \sqrt{2}f^\prime}} \\
\end{array}%
\right)\,, \nonumber \\
\langle L_2\rangle^T &=&  f^\prime\left(%
\begin{array}{ccc}
  -i\,\sin{\frac{\langle h\rangle}{\sqrt{2} f^\prime}}\,, &
  0 \,,&
  \cos{\frac{\langle h\rangle}{ \sqrt{2}f^\prime}} \\
\end{array}%
\right)\,, \eeqs where
 \beqs
f^\prime&=&\sqrt{\frac{f^2\,+\,12v^2_2\,+\,4v^2_1}{1\,+\,20v^2_2/f^2}}\,,
\nonumber \\
\langle |h| \rangle &=& \sqrt{2}f^\prime
\arctan{\sqrt{\frac{6v^2_2\,-\,3v^2_1}{f^2\,+\,6v^2_2\,+\,7v^2_1}}}\,.
\eeqs Appoximately we have
 \beqs
f^\prime&\approx& f \,,\nonumber \\
\langle |h| \rangle &=& \sqrt{12\,v^2_2\,-\,6\,v^2_1}\equiv v\,.
\label{eq:higgs-vev}
 \eeqs

Eqs.~(\ref{eq:higgs-mass}),~(\ref{eq:higgs-vev}) show a relation
between the Higgs doublet mass and its VEV as $m_h=v/\sqrt{3}$.
Using $v=246$~GeV, the Higgs boson mass is around 150~GeV.
Actually the ratio of the Higgs boson mass over $v$ depends on
order one Yukawa couplings in our model. Thus, the Higgs boson
mass can vary from 100~GeV to 200~GeV by choosing $y_6$ from $1/3$
to $2/3$.

After electroweak symmetry breaking, the lightest quark in the top
sector, which is identified as the top quark in the standard
model, has top Yukawa coupling \beqs y_t&\simeq&
\sqrt{\frac{y_1^2y_3^2}{2(y_1^2+y_3^2)}}\,. \eeqs Choosing the
same numerical values $y_1=y_3=2$ as in
Fig.~\ref{fig:triplet-potential}, $y_t\simeq 1$.

\section{Direct gauge mediation}

In our model, the soft masses of gauginos and sfermions arise from
gauge mediation. The messenger fields are embedded inside the SUSY
breaking sector, and therefore our model belongs to a direct gauge
mediation model.

The couplings in the superpotential in Eq.~(\ref{eq:sp})
explicitly break the continuous $U(1)_R$ symmetry to $R$ parity.
We calculate the one-loop Coleman-Weinberg potential of the pseudo-moduli ${\rm
Tr}\Phi_i$, and find that they develop  nonzero VEV's as $\langle {\rm
Tr}\Phi_i \rangle = O(\mu)$ (similar results can be found
in~\cite{R symmetry breaking}). In our model, it is crucial to
have non-zero VEV's of ${\rm Tr}\Phi_i$ to generate masses for
$SU(3)_W$ and $U(1)_X$ gauginos, which include winos and bino. But
the masses of gluinos can be generated independent of $\langle
{\rm Tr}\Phi_i \rangle$.

For the  gluino masses, $T_i$, $T_i^c$, $X_i$ and $X_i^c$, are the
``messengers" in our model. We only need to consider the last four
operators in Eq.~(\ref{eq:sp})
 \beqs
 \left(
   \begin{array}{cc}
     T_i & X_i \\
   \end{array}
 \right)
 \left(
   \begin{array}{cc}
     y^\prime_4\,\mu & y_3\,\mu \\
     y_3\,\mu & 0 \\
   \end{array}
 \right)
 \left(
   \begin{array}{c}
     T_i^c \\
     X_i^c \\
   \end{array}
 \right)\,,
  \eeqs
where $y^\prime_4\equiv y_4+ y_2\langle {\rm Tr}\Phi_i\rangle/\mu$.
The SUSY breaking mass terms of the messenger fields are given as $y_2\,F_{{\rm Tr}\Phi_i}(T_iT_i^c+h.c.)$. From the standard one-loop Feynman diagram calculation, we have the formula for the
masses of gluinos as
 \beqs
m_{\rm
gluino}&=&2\frac{g^2_3}{16\pi^2}\{\cos{\alpha_1}\cos{\alpha_2}[B(m_1,
M_1^2, M_2^2) \nonumber \\
&+&B(m_2, M_3^2, M_4^2)]+\sin{\alpha_1}\sin{\alpha_2}[B(m_1,
M_3^2, M_4^2) \nonumber \\
&+&B(m_2, M_1^2, M_2^2)]\} \,, \label{eq:gluino}
 \eeqs
where the factor of 2 comes from the presence of two ISS sectors
and the function $B(a, b^2, c^2)$ is defined as \beqs B(a, b^2,
c^2)&\equiv&
a\,\left(\frac{b^2\log{\frac{b^2}{a^2}}}{a^2-b^2}+\frac{c^2\log{\frac{c^2}{a^2}}}{c^2-a^2}\right)\,.
\nonumber \eeqs Here, $m_i$ are fermion masses and $M_i$ are
scalar masses. Corresponding expressions are
 \beqs
m_{1,2}&=&\frac{y^\prime_4\pm\sqrt{y^{\prime\,2}_4\,+\,4\,y_3^2}}{2}\,\mu \,,\nonumber \\
M^2_{1,3}&=&-\frac{F}{2}+\frac{m_1^2+m_2^2}{2} \nonumber \\
&&\pm\frac{\sqrt{F^2+(m_1^2-m_2^2)^2+2F(m_2^2-m_1^2)\cos{2\theta}}}{2}\,,
\nonumber \\
M^2_{2,4}&=&\frac{F}{2}+\frac{m_1^2+m_2^2}{2} \nonumber \\
&&\pm\frac{\sqrt{F^2+(m_1^2-m_2^2)^2+2F(m_1^2-m_2^2)\cos{2\theta}}}{2}\,,
\nonumber
 \eeqs
 with
 \beqs
\theta&\equiv&-\frac{1}{2}\arctan{\frac{2\,y_3}{y^\prime_4}}\,, \nonumber \\
\alpha_{1,2}&\equiv&\frac{1}{2}\arctan{\frac{\mp F
\sin{2\theta}}{m_2^2-m_1^2\pm F \cos{2\theta}}}\,, \nonumber \\
F&\equiv& y_2\,F_{{\rm Tr}\Phi_i}\,=\,y_2\,y\,\mu^2 \,.\nonumber
 \eeqs

Three interesting limits for Eq.~(\ref{eq:gluino}) can be studied:
\begin{itemize}
    \item When $F=0$ or $y_2=0$, we have $M_1=M_2=m_1$ and $M_3=M_4=m_2$.  As $B(a, a^2, a^2)=0$,
    the gluinos mass is zero. For a small $F$, expanding the gaugino mass in
terms of $F/m^2_i$, we find that the leading non-vanishing term
     is proportional to $F^3$ which agrees with~\cite{Izawa:1997gs}. In terms of the fermion masses, the leading term
     of gluino masses is
        \beqs
        &&-\frac{g^2_3}{8\pi^2}F^3\{[(m_1^2-m^2_2)(m_1^2-4m_1m_2+m_2^2)(m_1^6+m_1^4m_2^2 \nonumber \\
        &&+8m_1^3m_2^3+m_1^2m_2^4+m_2^6)+12m_1^4m_2^4(m_1^2+m_2^2) \nonumber \\
        &&\log{(m_1^2/m_2^2)}]/[6m_1^2m_2^2(m_1-m_2)^7(m_1+m_2)^4]\}\,.
        \eeqs

    \item When $y_3=0$, this is the simpliest case of gauge
    mediation models with only one messenger. In this case,
    $\theta=\alpha_1=\alpha_2=0$, $m_1=y_4\mu$, $m_2=0$,
    $M_{1,2}^2=m_1^2\mp F$ and $M^2_3=M^2_4=0$. After
    algebraic manipulations, the masses of the gluino are $m_{\rm gluino}=2g^2_3/16\pi^2\,\times F/m_1\times\,g(F/m^2_1)$,
    with $g(x)\equiv [(1+x)\log{(1+x)}+(1-x)\log{(1-x)}]/x^2$.
    This agrees with the result in the literature~\cite{Martin:1996zb}.

    \item When $y^\prime_4=0$, the continuous $U(1)_R$ symmetry is
    unbroken. We have $m_{1,2}=\pm y_3\mu$, $\theta=-\pi/4$ and
    $\alpha_1=\alpha_2=\pi/4$. Considering that $B(a, b^2,
    c^2)=B(-a, b^2, c^2)$, we have $m_{\rm gluino}=0$.
\end{itemize}

In our model, all the Yukawa couplings are generally order one numbers, so $F$ and $\mu^2$ are at the same scale. With $\mu=O(10~{\rm TeV})$ and $\sqrt{F}=O(10{\rm ~TeV})$, we have
 \beqs
 m_{\rm gluino}&=&2 \frac{g_3^2}{16\pi^2}O(\frac{F^3}{\mu^5})\,=\,O(200~{\rm{\rm GeV}})\,.
 \eeqs

For the $SU(3)_W$ gauginos, the mixed messengers are $H_i,
\tilde{H}_i, N_i, \tilde{N}_i$. Considering the non-zero VEV
$\langle {\rm Tr}\Phi \rangle$ and applying the same formula in
Eq.~(\ref{eq:gluino}) by substituting $y_4^\prime\rightarrow
y\langle {\rm Tr}\Phi \rangle/\mu$, $y_3\rightarrow y$ and
$F\rightarrow y\mu^2$, we have the masses of those $SU(3)_W$
gauginos to be also of $O(200~{\rm{\rm GeV}})$. Similar results
are derived for the $U(1)_X$ gaugino with two sets of mixed
messengers: one is $T_i$, $T_i^c$, $X_i$ and $X_i^c$; the other
one is $H_i, \tilde{H}_i, N_i, \tilde{N}_i$. The winos and bino
are parts of $SU(3)_W\times U(1)_X$ gauginos, and hence have
masses of $O(200~{\rm{\rm GeV}})$.

Finally, the masses of squarks and sleptons are generated through
the traditional two-loop diagrams and are also of $O(200~{\rm{\rm
GeV}})$.

\section{Mass Spectrum and Phenomenology}
SUSY and the global $U(4)^2$ symmetry are broken at order $10~{\rm
TeV}$. Most fields in the ISS sectors have masses at this scale.
Especially, the Higgsino mass is
\begin{equation}
m_{\rm Higgsino}=O(\mu) \sim{O(10~{\rm TeV})}\,.
\end{equation}
This arises naturally in our model and will not cause any
fine-tuning problem for the Higgs boson mass due to the little
Higgs mechanism present in the effective low energy theory. In
addition, we have heavy colored superfields $T_i,
T_i^c,X_i,X_i^c$, which are responsible for $SU(3)_W\times U(1)_X$
breaking to $SU(2)_w\times U(1)_Y$.

At the  $U(3)^2$ breaking scale $f =O(1~{\rm TeV})$, there are the
pesudo-moduli $\Phi_i$ and their fermionic partners containing
electroweak triplets and doublets, and massive vector superfields
$W^\prime$'s and $Z^\prime$ corresponding to $SU(3)_W\times
U(1)_X/SU(2)_w\times U(1)_Y$ . Colored superfields $p_i,p_i^c$
also have masses of this scale.

The gauginos and sfermions in the minimal supersymmetric standard
model obtain masses from direct gauge mediation as described in
the previous section and are of order a few hundered ${\rm GeV}$.
Finally we comment on the masses of the standard model singlets in
our model. Most singlets in the ISS sectors, $Z_i$ and $S_i$ have
masses at or above $1$~TeV. The singlet PNGB in the coset space
$U(1)_1\times U(1)_2/U(1)_X$ obtains a mass of $O(100~{\rm GeV})$
from the operator $y^\prime Z_3(\tilde{H}_1H_2-\mu^2)$ added to
achieve the vacuum alignment.

In summary, we list a sample of the mass spectrum of the particles
charged under the standard model gauge  group in
Table~{\ref{tab:spectrum}}.

\begin{table}[h!]
\begin{center}
\begin{tabular}{c|cl}
\hline \hline &\quad&  Higgsino\\
$\sim20~{\rm TeV}$ & \quad&  heavy colored superfields  $T,T^c,X,X^c$ \\
 &\quad &heavy doublets in scalars $K_i$ \\
 &\quad &superfields $N_i$ and $\tilde{N_i}$ \\
 &\quad & heavy singlets \\
 \hline
 & \quad&   triplets and doublets in  $\Phi_i$ \\ $\sim1~{\rm TeV}$& \quad&  vector superfields: $W^\prime$'s and $Z^\prime$  \\  & \quad&  colored superfields $p\,,p^c$  \\
 & \quad& singlets \\
\hline
$250\sim500~{\rm GeV}$ & \quad& gauginos: gluinos, winos and bino \\
& \quad&  sfermions: squarks and sleptons \\
\hline
$\sim100~{\rm GeV}$ &\quad&  the Higgs boson \\
&\quad&  the singlet PNGB \\
\hline\hline

\end{tabular}
\end{center}
\caption{A sample of the mass spectrum of the particles in our
model. This is only the order of magnitude estimate. The Higgs
boson mass can vary from 100~GeV to 200~GeV depending on the order
one Yukawa couplings in our model.}\label{tab:spectrum}
\end{table}

The scales in Table~\ref{tab:spectrum} are only an order-of-magnitude estimate.
The detailed spectrum depends on order one parameters in the model, and is subject
to constraints from electroweak precision measurements. To estimate the constraints,
we compare our model with the simplest little Higgs model.  Previous analyses on the simplest little Higgs model~\cite{constrain_sim} yield lower bounds for the masses of $W^\prime$'s, and then a lower bound for $\sqrt{f_1^2+f_2^2}\sim f$ about $3\sim 6$ TeV, depending on different choices of fermion charge assignments. Here, $f_1$ and $f_2$ are the magnitudes of the two $U(3)$ breaking VEV's (we have assumed $f_1=f_2=f$ for the sake of simplicity throughout the paper, but this is not essential for our model). As discussed below Eq.~(\ref{eq:higgs-vev}), the Higgs mass remains variable when the scale of $f$ is fixed. Therefore, we can achieve $f\gtrsim 3$ TeV without introducing significant fine tuning. At and below the TeV scale,
in addition to particles present in the simplest little
Higgs model, there are extra scalars and superparticles from the SUSY breaking sector.
The TeV scale scalars contribute to electroweak observables only at loop levels unless the $SU(2)_w$ triplets develop VEV's to break the custodial symmmetry~\cite{Chen:2006pb}, which is not the case in our model. The superparticles contribute to the electroweak observables only at loop levels. The full analyses of their effects again depend on order one parameters and are beyond the interests of this paper.

A lot of particles in our model are at or below the TeV scale and thus can be produced at the upcoming LHC. The spectrum bears a resemblance of the super-little Higgs models, namely, there exist heavy gauge bosons and fermions predicted by the little Higgs mechanism, and superpartners of the SM fields predicted by SUSY. Indeed, our model can be viewd as a UV completion of the super-simplest little Higgs model.  However, there are also important differences between our model and previous super-little Higgs models.  First, as discussed above, the Higgsino is necessarily absent from TeV scale spectrum.  Second, our model predicts  extra scalars from the SUSY breaking sector.

\section{Conclusions}

The SUSY breaking sector is usually treated as a ``hidden" sector
and  its effect is only mediated to the standard model through
messenger fields. In this paper, we have explored another approach
which is to identify the meta-stable SUSY breaking sector as an
extended Higgs sector. In our model, the standard model Higgs
doublet emerges from the meta-stable SUSY breaking sector as a
PNGB, and the electroweak scale is naturally two-loop factor below
the SUSY breaking scale.

We have employed two identical meta-stable SUSY breaking sectors
with $U(4)^2$ global symmetry. At order 10 TeV, SUSY is broken and
the global symmetry is broken to $U(3)^2$. Most fields in the SUSY
breaking sectors have tree-level masses of order 10 TeV. Among
those massless Nambu-Goldstone boson fields, there are two
triplets $L_1$ and $L_2$, in which the standard model Higgs boson
sits.

The electroweak gauge group, $SU(2)_w\times U(1)_Y$, is extended
to a $SU(3)_W\times U(1)_X$ gauge group embedded in the unbroken
global symmetry. The gauge interaction explicitly breaks the
global $U(3)^2$ symmetry  and makes the triplets $L_1$ and $L_2$
PNGB's. Introducing additional vector-like quarks in the top
sector, we have studied the one-loop effective potential of $L_1$
and $L_2$ and found that $L_1$ and $L_2$ obtain negative masses.
With order one quartic potential, $U(3)^2$ spontaneously breaks to
$U(2)^2$ at the scale of 1 TeV.  At the same time the gauge
symmetry $SU(3)_W\times U(1)_X$ is broken to $SU(2)_w\times
U(1)_Y$ and a light doublet PNGB is identified as the Higgs
doublet. At order 1 TeV, aside from the massive vector superfields
$W^\prime$'s and $Z^\prime$ and additional colored superfields, there
are pseudo-moduli fields including electroweak triplets and
doublets from the dynamical SUSY breaking sectors. The soft masses
of gauginos and sfermions in the minimal supersymmetric standard model  arise at one loop and two loops, respectively, in a similar way as in gauge mediation, and are of order a few hundred GeV. It is interesting that the Higgs sector plays the role of a messenger field, which distinguishes our approach from traditional gauge mediation models.

The Higgs doublet, which is a PNGB inside $L_1$ and $L_2$, has a
negative mass with another loop factor below the triplet masses
and triggers the electroweak symmetry breaking.  With order one
quartic potential generated through the sliding singlet mechanism,
the correct Higgs doublet VEV can be obtained and the Higgs boson
mass can vary from 100 GeV to 200 GeV depending on order one
Yukawa couplings in our model.

 \acknowledgements

We thank Witold Skiba for advice and help throughout this work. We
also thank Thomas Applequist, Maurizio Piai and John Terning for useful
discussions.

The research of Y.B. and J.F. is partially supported by Department
of Energy grant No. DE-FG02-92ER-40704. The work of Z.H. is
supported in part by the US Department of Energy grant No. DE-FG03-91ER-40674, and by the UC Davis HEFTI program.


\end{document}